\documentclass[sigconf,authorversion,natbib]{acmart}

\renewcommand\footnotetextcopyrightpermission[1]{} % removes footnote with conference information in first column
\newcommand\blfootnote[1]{%
  \begingroup
  \renewcommand\thefootnote{}\footnote{#1}%
  \addtocounter{footnote}{-1}%
  \endgroup
}

\usepackage{booktabs} % For formal tables
\usepackage{fancyhdr}
\usepackage{url}
\usepackage{color}

\usepackage{amsmath}
\usepackage{graphicx}
\usepackage{amssymb}
\usepackage{graphicx}
\usepackage{multirow}
\usepackage{array}
\usepackage{multirow}
\usepackage{booktabs}
\usepackage{floatflt}
\usepackage{caption}
\usepackage{subcaption}
\usepackage{mwe}

\usepackage{multirow}
\usepackage{rotating}
\setcitestyle{numbers,sort&compress}
\setcopyright{none}
\acmDOI{10.475/xxx}

\usepackage{bbm}
\setcitestyle{numbers,sort&compress}
\acmConference[ICTIR'17]{ICTIR}{2017}{Amsterdam, NL}
\acmYear{2017}
\copyrightyear{2017}

\acmPrice{xxx}

\begin{document}
\begin{sloppypar} 

\title{DSRIM: A Deep Neural Information Retrieval Model Enhanced by a Knowledge Resource Driven Representation of Documents} %\titlenote{A Knowledge resource-driven Representation of Text for Deep IR}
%\titlenote{DSRIM: Deep Semantic Resource Inference Model}
%\subtitlenote{The full version of the author's guide is available as
 % \texttt{acmart.pdf} document}

%\authornote{Dr.~Trovato insisted his name be first.}
%\affiliation{%
%  \institution{Institute for Clarity in Documentation}
%  \streetaddress{P.O. Box 1212}

\author{Gia-Hung Nguyen}
\affiliation{%
 \institution{IRIT, Universit\'e de Toulouse, CNRS }
 \streetaddress{Toulouse, France}
}
 \email{gia-hung.nguyen@irit.fr}
 
\author{Laure Soulier}
\affiliation{%
 \institution{Sorbonne Universit\'es, UPMC Univ Paris 06 CNRS, LIP6}
 \streetaddress{UMR 7606, 4 place Jussieu 75005 Paris, France}
} 
\email{laure.soulier@lip6.fr}
 
\author{Lynda Tamine}
\affiliation{%
 \institution{IRIT, Universit\'e de Toulouse, CNRS }
 \streetaddress{Toulouse, France}
} 
\email{tamine@irit.fr}
 
\author{Nathalie Bricon-Souf}
\affiliation{%
 \institution{IRIT, Universit\'e de Toulouse, CNRS }
 \streetaddress{Toulouse, France}
}
\email{nathalie.souf@irit.fr}

\renewcommand{\shortauthors}{GH Nguyen, Laure Soulier, Lynda Tamine and Nathalie Souf.}

\begin{abstract}
The state-of-the-art solutions  to the vocabulary mismatch  in information retrieval (IR)   mainly aim at    leveraging either the relational semantics provided by external resources or the distributional semantics,  recently  investigated by deep neural approaches. 
Guided by the intuition that the relational semantics might improve the effectiveness of deep neural approaches, we propose the Deep Semantic Resource Inference Model (DSRIM) that relies on: 1) a representation of raw-data that models the relational semantics of text by jointly considering objects and relations expressed in a knowledge resource, and 2) an end-to-end neural architecture that learns the query-document relevance by leveraging the distributional and relational semantics of documents and queries.
The experimental evaluation carried out on two TREC  datasets from TREC Terabyte and TREC CDS tracks relying respectively on WordNet and MeSH resources, indicates that our model outperforms state-of-the-art semantic and deep neural IR models.
\end{abstract}




\maketitle
\section{Introduction}
\blfootnote{This is the author's pre-print version of the work. It is posted here for your personal use, not for redistribution. Please cite the definitive version which will be published in Proceedings of ICTIR 2017}

Tackling the vocabulary mismatch has been a long-standing and major goal in information retrieval (IR). 
To infer and match discrete word senses within the context of documents and queries being matched, one line of work makes use of hand-labeled external knowledge resources such as linguistic resources and knowledge graphs. 
In IR, such resources allow to exploit objects and their relations (e.g., synonymy, hyperonymy) within, e.g., query or document expansion \cite{agirre2010document,xiong2015query} to lower the vocabulary mismatch between queries and documents; this is referred  to as the \textit{relational semantics}. 

Another line of work attempts to automatically infer hidden word senses from corpora using word collocations by performing dimensionality reduction techniques, such as latent semantic indexing \cite{Deerwester90indexingby} or, more recently, representation learning  \cite{mikolov:w2v,Pennington:2014} leading to \textit{distributional semantics}.  The latter was successfully exploited within deep neural networks for supporting search tasks \cite{guo2016deep,huang:dssm,severyn:shorttextrank}.  
One of the first contributions in the field relies on siamese architectures, opposing queries and documents in a two-branch network \cite{huang:dssm,severyn:shorttextrank}. However, these architectures are not yet mature since the  learning of a relevance function suffers from several limitations: 1) tackling traditional IR models (e.g.,  BM25 or language models) remains a difficult task \cite{guo2016deep,huang:dssm,hu2014convolutional} and 2) learning the relevance function on full text does not allow the network convergence, even though evaluated on  search logs of commercial search engines, leading to focus on a query-document title matching  \cite{guo2016deep,huang:dssm,severyn:shorttextrank}. 

Guided by the intuition that the relational semantics could complement distributional semantics and the motivation that siamese networks are under-explored in IR \cite{nguyen2016toward}, we investigate how to leverage both knowledge resources and siamese architecture to perform ad-hoc IR.
Specifically, we first model in a low-dimensional vector the relational semantics of text by jointly considering objects and relations expressed in knowledge resources. Then, we investigate the hypothesis that combining the distributional and the relational representations of text would enhance its representation for a ranking task. 
To the best of our knowledge, this is one of the first approach combining distributional and relational semantics in a  neural architecture with the goal to enhance  document-query ranking. 
More particularly, our contributions are twofold:

\noindent $\bullet$ A {\em Deep Semantic Resource Inference Model (DSRIM)}   leveraging:

- An input raw level representation of queries and documents relying on a  {\em knowledge resource-driven representation}.  More particularly, the premise of the latter representation relies on assumptions that a text is a bag of identified objects from a knowledge resource, and that semantically similar texts are deemed to entail similar/related objects.   
To deal with a large number of object-to-object relations, we propose the \textit{relation mapping} method that aims at projecting pairs in a low-dimensional space of object clusters. Our method is flexible since it can be used with any resource providing objects and relations between objects.

-  An end-to-end siamese neural network which learns an enhanced document-ranking function using input vectors combining both the distributional and the knowledge resource-driven representations of document/query. 

\noindent $\bullet$ A {\em thorough experimental evaluation} aiming at assessing the quality of the knowledge resource-driven representation and the effectiveness of  DSRIM. We use two TREC datasets, namely TREC PubMed CDS and TREC GOV2 Terabyte, and two knowledge resources, respectively MeSH\footnote{https://www.nlm.nih.gov/mesh/} and WordNet\footnote{http://wordnet.princeton.edu}. It is worth mentioning that, unlike previous work \cite{huang:dssm,severyn:shorttextrank} experimentally evaluated on document titles, our experiments are performed using full-texts.  

The rest of this paper is organized as follows. After reviewing the related work in Section 2, we motivate and then describe the DSRIM model in Section 3. Section 4 details the experimental protocol. 
 Section 5 discusses the experimental results. Section 6 concludes the paper and introduces perspectives.

% ======================
% Section Related work
% ======================
\vspace{-0,1cm}
\section{Related work}
\subsection{On the Semantic Representation of  Words, Documents, Objects, and Relations.}
The potential of word semantic representations  learned through neural approaches has been introduced in \cite{mikolov:w2v,Pennington:2014}, opening several  perspectives in  NLP and IR tasks.  Indeed, several work focuses on the representation of sentences \cite{Mikolov:phrases}, documents \cite{Mikolov:documents,Vulic:2015}, and also objects and relations expressed in knowledge resources \cite{Bordes:2013,Ji:2015,McCallum:2015,Wang:2014}.  Within the latter, the main principle consists in learning the representation of objects and relations on the basis of  object-relation-object triplets, relying on the assumption that the embedding of object $o_i$ should be close to the embedding translation of object $o_j$ by relation $r$, namely $o_i \simeq o_j+r$ (TransE model) \cite{Bordes:2013}.   Extensions have been proposed considering, e.g.,  different object representations according to the semantic relation type (TransH) \cite{Wang:2014} or  a dynamic mapping between objects and relations constrained by their diversity (TransD) \cite{Ji:2015}.
Moreover, knowledge resources have been used to enhance the distributed representation of words for representing their underlying concepts \cite{Faruqui:retrofitting,Liu:2016,xu:rc-net,yu:RCM}. 
Faruqui et al. \cite{Faruqui:retrofitting} propose a ``retrofitting" technique consisting in a leveraging of lexicon-derived relational information, namely adjacent words of concepts, to refine their associated word embeddings.  
Other  work  \cite{xu:rc-net,yu:RCM}  introduces  an  end-to-end  approach that rather adjusts the objective function of the neural language model by either leveraging the relational and categorical knowledge to learn a higher quality word embeddings (RC-NET model) \cite{xu:rc-net} or extending  the  CBOW  model using prior relational knowledge \cite{yu:RCM}.

\subsection{On using Knowledge Resources in IR.}
Both general/specific linguistic bases (e.g., WordNet or UMLS respectively) and large-scale knowledge graphs (e.g., Freebase) represent external resources that offer valuable information about word semantics through objects (e.g., words, entities, or concepts) and their associated relations (e.g., ``is-a", ``part-of"). Based on the use of such resources, a first line of work in IR aims at increasing the likelihood of term overlap between queries and documents through query expansion \cite{xiong2015query,pal2014improving} or document expansion \cite{agirre2010document}. Among models expanding queries, Xiong et al. \cite{xiong2015query} propose two algorithms relying on the category of terms in FreeBase. While the unsupervised approach estimates the similarity between the category distribution of terms in documents and queries, the supervised approach exploits the ground truth to estimate the influence of terms. Authors in \cite{pal2014improving} propose a query expansion technique using terms extracted from multiple sources of information. For each query term, candidate expansion terms in top retrieved documents are ranked by combining their importance in pseudo-relevant documents and their semantic similarity based on their definition in WordNet. 
Unlikely, Agirre et al. \cite{agirre2010document} propose a document expansion technique based on the use of a random walk algorithm identifying from WordNet the most related concepts. 
The second line of work leverages relations modeled in knowledge resources at the document ranking level \cite{xiong2015b}. For instance, authors in \cite{xiong2015b} propose a learning-to-rank algorithm based on  objects of knowledge resources that are related to a given pair of query-document. 

\vspace{-0,1cm}
\subsection{On using Deep Neural Networks in IR.}
Recently, a large amount of  work has shown that deep learning approaches are highly efficient in several IR tasks (e.g., text matching \cite{huang:dssm}, question-answering \cite{Bordes:2014}).  
A first category of work uses neural  models for IR tasks \cite{Ai:2016,mitra:dual,Zamani:2016} to integrate  embeddings in IR relevance functions.
The second category of work, closer to our contribution, consists in end-to-end scoring models that learn the relevance of document-query pairs via latent semantic features \cite{huang:dssm,hu2014convolutional}. For instance, the Deep Semantic Structured Model (DSSM) \cite{huang:dssm} applies a siamese deep feed-forward network on document and query representations obtained by a word hashing method. The network aims at learning their latent representations and then measuring their relevance score using a cosine similarity.
As an extension of the DSSM, Shen et al. \cite{shen:clsm} use a convolutional-pooling structure, called Convolutional Latent Semantic Model (CLSM). In the same mind, Severyn and Moschitti \cite{severyn:shorttextrank} apply a convolution  to learn the optimal representation of short text pairs as well as the similarity function. However, these model parameters are hard to learn, which leads authors to only consider the matching between query-title pairs. Guo et al. \cite{guo2016deep} also argue that tackling traditional IR models (e.g.,  BM25 or language models) remains a difficult task.   To bypass this limitation, another line of work \cite{lu2013deep, guo2016deep,Pang:2016} rather aims at building a local interaction map between inputs, and then uses deep neural networks to learn hierarchical interaction patterns. The DeepMatch model \cite{lu2013deep} integrates a topic model into a fully connected deep network based on the word interaction matrix while the MatchPyramid model \cite{Pang:2016} applies a convolution to an interaction matrix estimated on the basis of word representation. Guided by the intuition that interaction matrix is more appropriate for the global matching and lacks of the term importance consideration,  authors in  \cite{guo2016deep} model local interactions of query-document terms through  occurrence histograms.

\begin{figure*}[t]
\centering
\includegraphics[width=0.8\textwidth]{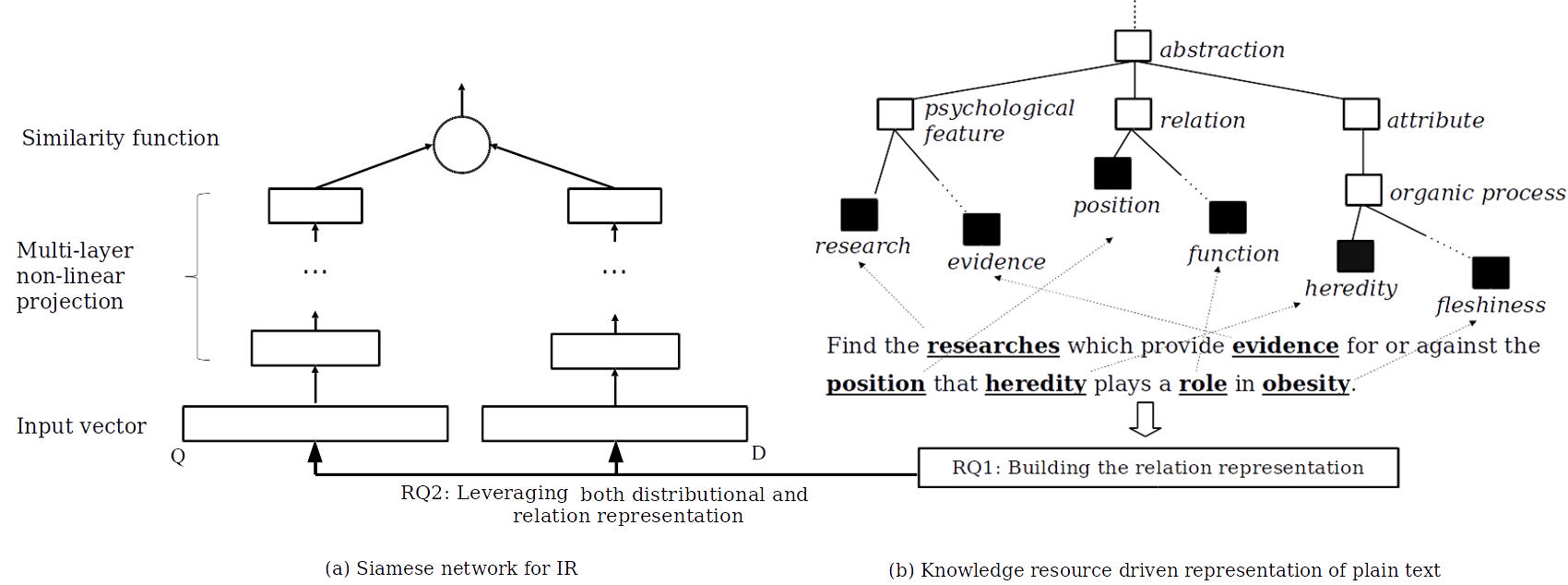}
\vspace{-0.4cm}
\caption{General issue of our contribution and research questions}
\label{fig:motivation}
\vspace{-0.2cm}
\end{figure*}

\begin{figure*}
\centering
\includegraphics[width=0.8\textwidth,keepaspectratio]{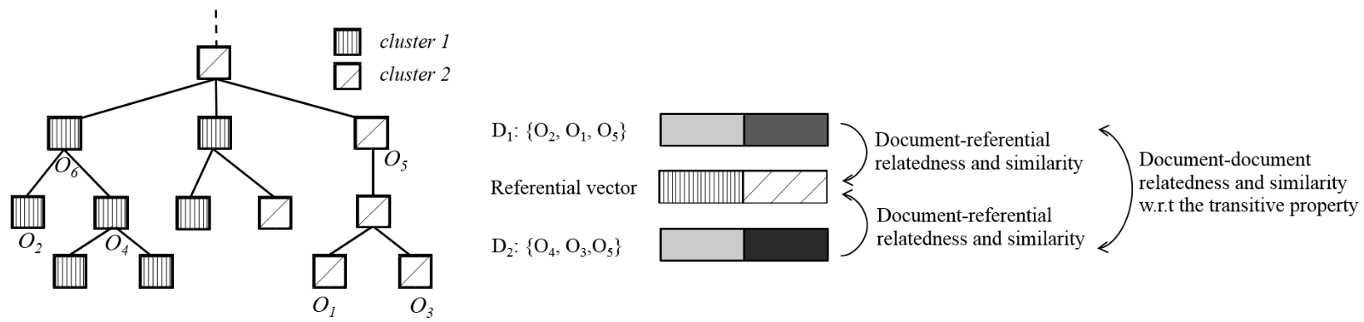}
\vspace{-0.4cm}
\caption{Intuition of the transitive property in knowledge resource-driven representation of documents}
\vspace{-0.2cm}
\label{fig:transitive}
\end{figure*}
%%Laure\vspace{-0.4cm}
\section{DSRIM: Deep Semantic Resource Inference Model}
\subsection{Motivation}
\label{problem}
The literature review highlights that: 1) plain text and knowledge resources are complementary for both learning distributional representations and enhancing IR effectiveness  \cite{agirre2010document,pal2014improving,xiong2015query}, and that 2) neural approaches in IR, and more particularly siamese architectures, have a great potential for ad-hoc search but could still be improved to compete with traditional IR models \cite{guo2016deep}. In this contribution, we address the problem of bridging the semantic gap in IR by leveraging both deep learning approaches \cite{huang:dssm,severyn:shorttextrank} and valid knowledge expressed in knowledge resources  \cite{xiong2015query,pal2014improving}. 
In contrast to previous work in deep IR models \cite{huang:dssm,hu2014convolutional,Pang:2016,severyn:shorttextrank} relying only on the distributional semantics of texts and work on the semantic representation of objects and relations only leveraging  knowledge resources \cite{Bordes:2013,Faruqui:retrofitting,Wang:2014,xu:rc-net}, our main concern is to estimate a relevance function leveraging a semantic representation of documents that simultaneously takes into consideration objects and  their pairwise relations expressed in a knowledge resource.
With this in mind, we investigate the potential of siamese neural architectures, such as DSSM \cite{huang:dssm}, on full-text retrieval.  In this paper, we specifically address the two main following research questions as illustrated in Figure 1:

%\vspace{-0.2cm}
\begin{itemize}
\item  \textbf{RQ1}: How to model the relational semantics of texts at the raw data level by jointly leveraging objects and their relations expressed in knowledge resources?
\item  \textbf{RQ2}: How to learn  the query-document relevance function by combining the relational and distributional semantics of text in a siamese neural architecture?
\end{itemize}
Below, we detail our contributions w.r.t. each research question.

\vspace{-0.2cm}
\subsection{Knowledge Resource driven  Representation}
\label{KBrep}
Our aim is to model a text representation  that conveys their semantics with respect to a knowledge resource. 
The premise of this representation relies on two assumptions: 
(A1) a text is a bag of identified objects from a knowledge resource, and  (A2) semantically similar texts are deemed to entail similar and related objects.

Formally, a knowledge resource is built upon a relational graph $G=\left(V,E\right)$ where $V$ is a  node set  and $E$ is a edge set. Each node $v_i=<o_i, desc_i,>$ includes an object $o_i$ (e.g., word, entity) and its textual label $desc_i$ (e.g., preferred entry).  Each object $o_i$ is associated to a distributional representation $x_i^{d}$ (e.g.,  its  \textit{ParagraphVector} \cite{Ai:2016} obtained on the basis of its textual labels $desc_i$).  Each edge $e_{i,i'}$ expresses a semantic relation between  objects $o_i$ and $o_{i'}$.
We suppose that given the set $O$ of objects in the knowledge resource $G$, we can identify, for each text $T$, a set $O(T) \subset O$ of objects $o$.

While assumption A1 is easy to formalize through a binary vector modeling objects $o_i \in V$ or a vector  combining their distributional representation $x_i^{d}$, it does not allow to fulfill assumption A2. To cope with this issue, the perspective of a vector representing object-object pairs could be a good option to simultaneously capture: 1) the objects belonging to a text and 2) their similarity as well as their relatedness. However, the large number of potential pairwise objects, or more precisely object-to-object relations, in a knowledge resource would lead to a high dimensional and sparse vector. To face this issue, we propose the \textit{relation mapping method}, that: 1)  similarly to the word hashing method \cite{huang:dssm}, aims at reducing the dimensionality and the sparsity of the vector representation to make it scalable, and 2) allows  building representations of  both objects belonging to text $T$ and their relations according to assumption A2. We describe below our approach for achieving these two sub-goals.

$\bullet$ Sub-goal 1) \textit{Text representation vector space:} A naive approach consists in considering  objects from the knowledge resource as unit vectors of a $ |V| $-dimensional space. Even if the number of objects in the resource is significantly lower than the number of object-to-object relations, the scalability of the underlying framework remains questionable. To fit with sub-goal 1) and lower the dimensionality of the vectorial representation space, we rather consider clusters of objects as representative of each dimension of the vectorial space.  Assuming that object-to-object relations might express topical relatedness between objects, we propose to build $k$ topical clusters $c_j$ of objects $o_i \in O$ assumed to be mutually independent. The latter refers to the referential $\mathcal{R}=\{c_1,\dots,c_k\}$ of the knowledge resource.  In practice, we use the k-means clustering algorithm on the topical representation of objects, where the number of topical clusters $k$ would be experimentally tuned (see Section \ref{analysisVector}). Thus, we consider a k-dimensional space, in which $k$ is the number of topical clusters of objects.

$\bullet$  Sub-goal 2) \textit{Knowledge resource-driven text representation:} The representation $x^{KR}$ of text $T$   is a  k-dimensional  vector  $x^{KR} = (x^{KR}_1, ...,x^{KR}_k)$. To fulfill  sub-goal 2, our intuition is that two documents are likely to be similar if they mention objects that are gathered around the same topical clusters. Naturally, the degree of similarity between those documents would depend on the average relatedness and similarity of their objects with each object in the topical clusters $c_j$ of the referential $\mathcal{R}$.  This refers to as a transitive property, illustrated in Figure \ref{fig:transitive}.   Each document $D_1$ and $D_2$ is modeled through a 2-dimensional vector in which each element represents a topical cluster. The gray levels in the document representation express the relatedness and similarity degree of document objects with respect to the topical clusters. Although documents $D_1$ and $D_2$ are not characterized by the same objects, they  are as close to the referential, and accordingly, have similar representations. 
We compute  each element $x^{KR}_j$  as a combination of the importance $w_j ^{T}$ of  topical cluster $c_j$ given text $T$ and the relatedness $S_{relat}(c_j,O(T))$ of  objects $O(T)$  belonging to text $T$ w.r.t. topical cluster $c_j$:
%\vspace{-0.1cm}
\begin{equation}
x^{KR}_j = w_j ^{T}* S_{relat}(c_j,O(T))
\end{equation}

\subsubsection{Topical cluster importance score}
The  importance score $w_j^T$ of topical cluster $c_j$ expresses to what extent the set $O(T)$ of objects belonging to text $T$ are topically similar to objects belonging to topical cluster $c_j$. 
Intuitively,  the more topically similar the objects mentioned in the representations of texts $T$ and $T'$ with respect to the topical clusters, the more similar texts $T$ and $T'$. 
Assuming that objects belonging to a text represent a topical cluster, we rely on previous work dealing with clustering similarity \cite{clustering1} suggesting to   estimate the similarity between two sets of objects by aggregating similarities  between objects of these two different sets. More formally,  the topical cluster importance score between topical cluster $c_j$ and  object set $O(T)$  is estimated as:
\begin{equation}
w_j^T =Agg\_Function_{(o_m,o_n) \in O(T) \times c_j} sim_t(o_m,o_n) 
\end{equation}

where $Agg\_Function$ expresses an aggregation function (we consider here the maximum to capture the best topical similarity between objects);  $sim_t$  estimates the topical similarity between vector representations of objects (here, the cosine similarity between the vectorial representations of object textual descriptions).

\subsubsection{Topical cluster-text relatedness score}
The topical cluster-text relatedness score $S_{relat}(c_j,O(T))$ measures to what extent objects $o_i \in O(T)$ belonging to text $T$ are related to those of topical cluster $c_j$. Our intuition is that if the objects mentioned in texts $T$ and $T'$  are related to the representative of the same topical clusters, texts $T$ and $T'$ are more likely to be similar. Having in mind that state-of-the-art relatedness measures \cite{Pedersen2007} rely on the  computation of paths between objects, a scalable way allowing to measure this score is to consider the relatedness of objects $O(T)$ with respect to a representative object $R(c_j)$ of topical cluster $c_j$  (e.g., the most frequent object in the collection among objects belonging to topical cluster $c_j$). The impact of the method used for identifying the  representative is experimentally investigated (see Section 6.1). 
More formally, given a representative object $R(c_j)$ of topical cluster $c_j$, the topical cluster-text relatedness  $S_{relat}(c_j,O(T))$  estimates  the path length between object $R(c_j)$ and the object set $O(T)$:
\begin{equation}
\label{eq:sem_vector}
 S_{relat}(c_j,O(T)) =\sum_{o_m \in O(T)}\log{(1+sim_r(R(c_j),o_{m}))} \cdot \frac{avg\_no}{|O(T)|} 
\end{equation}

where $o_m$ is an object of the object set $O(T)$ characterizing text~$T$. $sim_r$ is a relatedness measure between objects (here the Leacock measure \cite{leacock1998combining}); $avg\_no$ is the average number of objects by document in the collection.   The normalization factor $\frac{avg\_no}{|O(T)|}$  avoids bias due to differences in text lengths in terms of the  number of objects.

\subsection{Model Architecture}

\subsubsection{Input.} \label{subsec:input}
We propose to characterize each text $T$ (whether extracted from a document or a query)  by an input vector $x_{input}= (x^{t}, x^{KR})$  modeled as a vector composed of two parts:

$\bullet$ \textit{Plain text representation $x^{t}$.} This feature represents  words of  full text $T$. Based on previous findings highlighting the effectiveness of distributed semantic representations to tackle the issue of large vocabulary in IR, we use the \textit{ParagraphVector} model \cite{Ai:2016}.

$\bullet$ \textit{Knowledge resource-driven representation $x^{KR}$.} This feature expresses the objects belonging to  text $T$ and their semantic relations expressed in the knowledge resource. This representation is built upon the \textit{relation mapping} method (see Section \ref{KBrep}).

\subsubsection{Learning the latent representation.}
For each sub-network branch, the input vector $x_{input}$ of text $T$ is projected into a latent space by means of $L$ hidden layers $l_i$ ($i=1,\cdots,L$) so as to obtain a latent semantic vector $y$ combining the distributional and relational semantics of text $T$. Each hidden layer $l_i$ and the latent semantic vector $y$ are respectively obtained by  non-linear transformations: 
 \begin{gather}
 	l_0 = x_{input} \nonumber \\
 l_i = f(W_{i-1} \cdot l_{i-1} + b_{i-1}) ~~~~~~ i=1,...,L\\
 y = f(W_L \cdot l_L + b_L) \nonumber
 \end{gather}
where $W_i$ and $b_i$ are respectively the weight matrix and bias term of the $i^{th}$ layer. The activation function $f(x)$ performs a non-linear transformation, namely the ReLU: $f(x) = max(0,x)$, which has been commonly used in deep learning works \cite{lecun2015deep}. The use of the ReLU function is motivated by the fact that it does not saturate to 1 when $x$ is high in contrast to the hyperbolic tangent \cite{huang:dssm}, avoiding to face to the gradient vanishing problem.

The latent semantic vectors $y_D$ and $y_Q$ of document $D$ and query $Q$ obtained through the non-linear transformations are used to estimate the document-query cosine similarity score $R(D|Q)$.% is estimated between vectors $y_D$ and $y_Q$.

\subsubsection{Loss function.}
Since   retrieval tasks refer to a ranking problem, we optimize the parameters of the neural network using a pairwise ranking loss based on the distance $\Delta$ of similarity between relevant document-query pairs, noted $(Q,D^{+})$, and irrelevant document-query pairs, noted $(Q,D_{p}^{-})$. Unlike \cite{huang:dssm}, it worth mentioning that we use the hinge loss function, more adapted for learning-to-rank tasks \cite{Chen:2009,dehghani2017neural}. To do so, we build a sample of document-query pairs in which we oppose, for the same query $Q$, one relevant document $D^{+}$ for $n$ irrelevant documents $D_p^{-},~ p =[1..n]$, as suggested in \cite{huang:dssm}.
The difference $\Delta$ between the similarity of the relevant pair $(Q,D^+)$ and the irrelevant ones $(Q,D_{p}^-)$ is defined as:
\begin{equation}
\Delta =  \bigg[ sim(Q,D^+) - \sum_{p=1}^{n} sim(Q,D_{p}^-)\bigg]
\label{eq:delta}
\end{equation}
where $sim(\bullet,\bullet)$ is the output of the neural network. Then, the DSRIM network is trained to maximize the similarity distance $\Delta$ using the hinge loss function $L$: $L = max(0, \alpha - \Delta)$ where $\alpha$ is the margin of $L$, depending on the $\Delta$ range.

% ======================
% Section 5: Dataset and Evaluation
% ======================
\section{Evaluation protocol}

\subsection{Datasets} 
We consider two datasets (statistics are presented in Table \ref{table:dataset}):

$\bullet$ The GOV2\footnote{http://ir.dcs.gla.ac.uk/test\_collections/gov2-summary.htm} dataset gathering .gov sites used in the TREC Terabyte campaign. We use topics from the 2004, 2005, and 2006 campaigns and  the narrative part of each topic as a query.

$\bullet$ The PMC OpenAccess\footnote{https://www.ncbi.nlm.nih.gov/pmc/tools/openftlist/} dataset with biomedical full-texts from PubMed used in the TREC-CDS campaign. The summaries of  topics of the 2014 and 2015 evaluation campaigns are used as queries.

\subsection{Implementation Details and Evaluation Methodology}
\label{sec:impl_detail}

To build the input layer, we pre-train a \textit{ParagraphVector} model on the plain text corpus for learning vector $x^t$. The vectors are sized to $100$, as suggested in \cite{Ai:2016}. 
The concepts used for building our knowledge resource-driven representation are extracted using appropriate tools, namely \textit{SenseRelate} on WordNet\footnote{http://wordnet.princeton.edu} resource \cite{pedersen2009wordnet} for the GOV2 dataset and Cxtractor\footnote{https://sourceforge.net/projects/cxtractor/} relying on \textit{MaxMatcher} \cite{Zhou:2006} applied on the 2015-MeSH version\footnote{https://www.nlm.nih.gov/mesh/} for the PMC dataset. We used for both the 'IS-A' relation.
For modeling the  representation $x^{KR}$, we tune two parameters: 1) the number of topical clusters:  we set the number $k$ of topical clusters to $k \in \{100,200\}$; 2) the choice of the representative object $\mathcal{R}(c_j)$ within each topical cluster:  we use three strategies: $idf_{min}$, namely the most frequent object; $idf_{max}$, the less frequent one; and $centroid$, the closest object to the  centroid. Concerning our model architecture, we set the number of hidden layers to 2 with a hidden vector size equals to 64 leading to an output layer of 32 nodes. Similarly to \cite{huang:dssm}, the number $n$ of irrelevant document-query pairs opposed to a relevant one is $4$ (Equation \ref{eq:delta}). Relevant/irrelevant document-query sets are randomly extracted from each dataset ground truth, supplying graded relevance judgments from 0 to 2 (relevance criteria: 1 and 2).

\begin{table}[t]
\centering
\caption{Statistics of the GOV2 and the PMC datasets}
\vspace{-0.4cm}
\label{table:dataset}
\begin{tabular}{p{5cm}cc}
\toprule
 & GOV2 & PMC \\ 
 \midrule
\# Documents & 25,000,000 & 733,138 \\ 
Average length of documents (\#words) & 1132.8 & 477.1 \\ 
\# Queries & 150 & 60 \\
\# Relevant pairs~~~ & 25,100~~~ & 8,346 \\
 \bottomrule
\end{tabular}
\vspace{-0.2cm}
\end{table}

To train our model parameters, we apply the 5-fold cross-validation method. The topics in each dataset are divided into 5 folds. For each fold retained as the test set for model evaluation, the other 4 folds are used to train and validate the model. The final retrieval results are averaged over the test results on 5 folds. The model is optimized using a 5-sample mini-batch stochastic gradient descent (SGD) regularized with a dropout equals to 0.3. Our model generally converges after 50 epochs over the training dataset.

For evaluating the ranking performance of our model and the different baselines, we perform a re-ranking  \cite{guo2016deep} which is carried out over the top 2,000 documents retrieved by the BM25 model on Lucene. Final results are estimated using the top 1000 documents of each re-ranking model according to the MAP metric.

\begin{table*}[t]
\centering
\caption{Cosine similarities of the knowledge resource-driven representation on most similar (Top\_10) and less similar (Less\_10)  documents, averaged on 100 random pivotal documents. diff: difference between Top\_10 and Less\_10}
\vspace{-0.4cm}
\label{tab:topflop}
\begin{tabular}{@{}lclllllll@{}}
\toprule
 	& \multicolumn{1}{l}{}      &         & \multicolumn{3}{c}{GOV2}           & \multicolumn{3}{c}{PMC}  \\ \midrule
\multicolumn{1}{l|}{\multirow{7}{*}{Clustering}}  & \multicolumn{1}{c|}{\#Clusters $k$}                 & \multicolumn{1}{l|}{Repres. obj. $R(c_i)$} & Top\_10 & Less\_10 & \multicolumn{1}{l|}{diff}   & Top\_10 & Less\_10 & diff \\ \cmidrule(l){2-9} 
\multicolumn{1}{l|}{}                             & \multicolumn{1}{c|}{\multirow{3}{*}{\#Cluster 100}} & \multicolumn{1}{l|}{idf\_max}    &    0.7490       &	    0.5776      &	\multicolumn{1}{l|}{0.1714}    & 0.5455  & 0.3035   & {0.2420}    \\
\multicolumn{1}{l|}{}                             & \multicolumn{1}{c|}{}                             & \multicolumn{1}{l|}{centroid}   &   0.7411   &	0.5693	&   \multicolumn{1}{l|}{0.1719}    & 0.4807  & 0.2862   & {0.1945}   \\
\multicolumn{1}{l|}{}                             & \multicolumn{1}{c|}{}                             & \multicolumn{1}{l|}{idf\_min}    &     0.7018	    &   0.5501  &  \multicolumn{1}{l|}{0.1518}    & 0.4975  & 0.2717   & {0.2259}   \\ \cmidrule(l){2-9} 
\multicolumn{1}{l|}{}                             & \multicolumn{1}{c|}{\multirow{3}{*}{\#Cluster 200}} & \multicolumn{1}{l|}{idf\_max}    &       0.7595    & 	0.5814     &   	\multicolumn{1}{l|}{0.1781}  & 0.6359  & 0.3885   & {0.2475}  \\
\multicolumn{1}{l|}{}                             & \multicolumn{1}{c|}{}                             & \multicolumn{1}{l|}{centroid}    &       0.7344	 &   0.5536	&   \multicolumn{1}{l|}{0.1808}  &{0.6464}  & {0.3842}   & 0.2621    \\  
\multicolumn{1}{l|}{}                             & \multicolumn{1}{c|}{}                             & \multicolumn{1}{l|}{idf\_min}   &    {0.7645}	&   {0.5660}	&   \multicolumn{1}{l|}{0.1985}  & 0.6485  & 0.4234   & 0.2251   \\ \midrule
\multicolumn{1}{l|}{\multirow{3}{*}{Baselines}} & \multicolumn{2}{c|}{Top$\_$concept, $k=200$}                                                       &        0.9034   &  	0.9013  &  	\multicolumn{1}{l|}{0.0021}   & 0.9861  & 0.9616   & 0.0245    \\
\multicolumn{1}{l|}{}                       & \multicolumn{2}{c|}{Top$\_$concept, $k= 100$}                                &      0.9123  &	0.9049  &	\multicolumn{1}{l|}{0.0074}   & 0.9817  & 0.9572   & 0.0245 \\ \cmidrule(l){2-9} 
\multicolumn{1}{l|}{}    &\multicolumn{2}{c|}{LDA}  &        0.4377   &  	0.3189  & 	\multicolumn{1}{l|}{0.1188}   & 0.2884  & 0.0518   & 0.2639    \\ %\cmidrule(l){2-9} 
%\multicolumn{1}{l|}{} &\multicolumn{2}{c|}{T-WE} &        0.8543   &  	0.7732  &  	\multicolumn{1}{l|}{0.0811}   & 0.3098  & 0.1947   & 0.1151    \\ 
\bottomrule
\end{tabular}
\vspace{-0.2cm}
\end{table*}

\subsection{Baselines}
To evaluate the quality of our knowledge resource-driven representation, we use two models building representations of documents:\\
$\bullet$ \textbf{Top$\_$concepts}: A naive version of our knowledge resource-driven representation  selecting the top $k$ frequent objects in the document collection as the representative objects ($k \in \{100,200\}$).  \\
$\bullet$ \textbf{LDA}: The well-known LDA topic model representing topic clusters from plain text \cite{Blei:2003} (vs. topical cluster relying on concepts and relations in DSRIM). \\

\vspace{-0.1cm}
\noindent To evaluate our model effectiveness, we use three types of baselines: \\
\textbf{1) Exact term matching models} to highlight the impact of both leveraging relational semantics and deep learning approaches:\\
$\bullet$ \textbf{BM25}: The well-known probabilistic model (\textit{BM25}). \\
$\bullet$ \textbf{LM-DI}: The language model based on Dirichlet smoothing  \cite{Zhai:2001}. \\
\textbf{2) Enhanced semantic matching models} to outline the impact of a deep neural model guided by knowledge resources for capturing text semantics:\\
\noindent $\bullet$ \textbf{LM-QE}: A language model applying a concept-based query expansion technique \cite{pal2014improving} in which candidate terms are ranked based on their similarity with descriptions in the knowledge resource. Default parameters mentioned in the paper are used. \\
$\bullet$ \textbf{LM-LDA}: The LM-LDA is a latent topical model using the language modeling framework \cite{wei2006lda}. \\
\textbf{3) Deep neural semantic matching models},  also based on a siamese architecture, to highlight the impact of combining relational and distributional semantics in neural approaches:\\
$\bullet$ \textbf{DSSM}: The state-of-the-art DSSM model \cite{huang:dssm}. We adopt the publicly released code\footnote{https://www.microsoft.com/en-us/research/project/dssm/} with  default parameter values. We  evaluate the DSSM on full-text documents.
\\
$\bullet$ \textbf{CLSM}: The DSSM extension  in which the feed-forward  network is replaced by a convolution. %for better capturing fine-grained contextual structures \cite{shen:clsm}. 
We also apply the publicly released CLSM code$^{8}$ on full-texts and use the default parameter values.\\

\vspace{-0.15cm}
\noindent To measure the impact of the different evidence sources taken into consideration for representing texts, we use three scenarios: \\ 
$\bullet$ \textbf{$DSRIM^{p2v}$}: Our proposed neural model based on an input representation of texts restricted to the plain text, namely $x^{t}$.\\
 $\bullet$ \textbf{$DSRIM^{kr}$}: Our proposed neural model based on our knowledge resource-driven representation of text, namely  $x^{KR}$.\\
$\bullet$ \textbf{$DSRIM^{kr+p2v}$}: Our proposed neural model based on an enhanced  representation of texts combining plain text representation $x^{t}$ and our knowledge resource-driven representation $x^{KR}$.

\begin{table*}[t]
\centering
\caption{Effectiveness comparison of baselines and DSRIM on GOV2 and PMC collections. \% Chg: Significant improvement/degradation of $DSRIM^{kr+p2v}$ (+/-). \textit{p-value}: Significance t-test: * : $0.01 < \alpha \leq 0.05$, ** : $0.001 < \alpha \leq 0.01$, *** : $\alpha \leq 0.001$
}
\vspace{-0.4cm}
\label{table:result}
\begin{tabular}{cc|lll|lll}
\hline
\multicolumn{1}{l}{} & \multicolumn{1}{l}{} & \multicolumn{3}{c|}{GOV2} & \multicolumn{3}{c}{PMC} \\ \hline
Model Type & Model & MAP & \%change & p-value & MAP & \%change & p-value \\ \hline
\multirow{2}{*}{\begin{tabular}[c]{@{}c@{}}Exact\\ Matching\end{tabular}} & BM25 & 0.1777 & +4.84 & 0.6691 & 0.0348 & -1.15 & 0.9628\\
& LM-DI & 0.1584 & +17.61 & 0.1644 & 0.0379 & -9.23 & 0.7109\\ 
\hline
\multirow{2}{*}{\begin{tabular}[c]{@{}c@{}}Semantic \\Matching\end{tabular}} 
& LM-QE & 0.0738 & +152.44 & 0.0001 *** & 0.0106  & +224.53 & 0.0008 *** \\
& LM-LDA & 0.0966 & +92.86 & 0.0001 ***  & 0.0185 &  +85.95 & 0.0323 *\\ 
\hline
\multirow{2}{*}{\begin{tabular}[c]{@{}c@{}}Deep \\ Matching \end{tabular}} 
& DSSM & 0.0418 & +345.69 & 0.0001 *** & 0.0095 &	+262.11 & 0.0008 *** \\
& CLSM & 0.0365 & +410.41 & 0.0001 *** & 0.0069 & +398.55 & 0.0001 *** \\ \hline
%\multirow{2}{*}{\begin{tabular}[c]{@{}c@{}}Document \\ representation \end{tabular}} 
%& DSRIM\textsuperscript{LDA} & 0.& & 0.& 0.&	& 0.\\
%& DSRIM\textsuperscript{T-WE} & 0.&& 0.& 0.& & 0.\\ \hline
\multirow{3}{*}{\begin{tabular}[c]{@{}c@{}}Our approach \end{tabular}} 
& DSRIM\textsuperscript{p2v} & 0.1115 & +67.09 & 0.0001 *** & 0.0183 &	+87.98 & 0.0460 *\\
& DSRIM\textsuperscript{kr} & 0.1801 & +3.44 & 0.7461 & 0.0307 &	+12.05 & 0.6829\\
& \textit{DSRIM\textsuperscript{kr+p2v}} &\textit{0.1863}& ~ & ~ &\textit{0.0344} & ~ & ~ \\ 
\hline
\end{tabular}
\vspace{-0.2cm}
\end{table*}

% ======================
% Section : Results
% ======================
\section{Results}
\subsection{Analyzing the Semantic Representation of Documents}
\label{analysisVector}

In this section, we propose to analyze our knowledge resource-driven representation through a  twofold objective: 1) identifying the optimal parameter setting of the vectorial representation and 2) assessing the validity of the built document vectors $x^{KR}$.

We assess the vectorial representation quality based on the intuition that semantically similar texts, modeled as bags of concepts, should have similar vectorial representations built following our approach; such representations should also discard non-similar documents \cite{mikolov:w2v,Mikolov:documents}. In practice, given a randomly selected  document (called a ``pivotal document"), a good vectorial representation should 1) ensure that the distance between the pivotal document and each other document of the collection is non-uniform, and 2) maximize the distance between its most similar documents and  its less similar ones. 
To this end:  1) we first identify for each given  pivotal document, the set $\mathcal{D}^{p}_{+}$ of its $10$ most semantically similar documents and the set $\mathcal{D}^{p}_{-}$ of the $10$ less semantically similar documents over the whole dataset using a concept-oriented metric proposed in \cite{corley2005measuring}, called in the remaining  the \textit{Corley} measure; and 2) then we compute the average cosine similarity of the representations of the pivotal documents with the sets $\mathcal{D}^{p}_{+}$ and  $\mathcal{D}^{p}_{-}$. Table \ref{tab:topflop} presents the comparative results  for 100 randomly selected pivotal documents and suggests the following statements: % used for building our knowledge resource-driven representation.

$\bullet$ The difference in terms of cosine value range between both datasets (higher for  GOV2) conjectures that representing texts using objects and relations expressed in a knowledge resource seems to be more difficult for the PMC dataset. This could be explained by the fact that this dataset focuses on a particular application domain (namely, the medical vs. general for GOV2) that might imply a more technical vocabulary.

$\bullet$ Regarding the method used for defining the vectorial representation space (sub-goal 1; Section 4.1), we can see that our proposed approach for identifying the referential based on the object clustering is more effective than  both baselines, respectively Top$\_$concept and LDA.  
Indeed, the similarity differences of both document sets $\mathcal{D}^{p}_{+}$ and  $\mathcal{D}^{p}_{-}$ obtained by the baselines are very small ($<0.11$ for both datasets, except LDA for PMC, vs. higher than $0.15$ for our clustering approach). 
It is worth to mention that the Top$\_$concept baseline particularly fails to discriminate between the most/less similar documents for both datasets given the high values of  cosine values ($>0.90$). Also, the small cosine values obtained using the LDA baseline for the most similar documents ($<0.5$) show that the LDA representation is not able to build close document representations. In contrast, we outline that cosine values for our clustering approach seem to be more intuitive, with an average cosine for the GOV2 dataset  higher  than $0.6$   for the most similar documents and lower than $0.6$  for the less similar ones (respectively $0.5$ for the PMC dataset). These statements suggest that our referential building approach based on topical clustering seems reasonable.

$\bullet$ Focusing on the methods used for the knowledge resource-driven representation (sub-goal2; Section 4.1) and more particularly, the one used for choosing the topical cluster representative, we can notice that the average similarities between  pivotal documents and the set of top similar ones are more important for a higher number of clusters (e.g., up to 0.6485 for $k=200$  vs. 0.5455 for $k=100$ for the PMC dataset). Also, this setting allows obtaining higher differences between the most vs. less similar documents (with at least $0.2251$ vs. $0.1945$ for respectively $k=200$ and $k=100$ for the PMC dataset, and $0.1781$ vs. $0.1518$ for the GOV2 dataset). These results highlight the importance of achieving a reasonable ratio between the knowledge resource size (in terms of the number of object-object-relations) and the number of representative clusters of objects to better capture the semantic representation of documents.
With this in mind, the best scenario for $k=200$ allowing to distinguish the most vs. the less similar documents consists in selecting the  closest  object  to the  cluster centroid ($centroid$) as the representative object for the PMC dataset while these are no significant differences between the three methods for the GOV2 dataset. Given that the centroid method is more intuitive with the assumptions used  for building the referential, we retain the setting with $200$ topical clusters and the $centroid$ method for encoding the representative object.
\vspace{-0.2cm}
 \subsection{Measuring the Model Effectiveness}
We present here the performance of our model on both datasets GOV2 and PMC. Table \ref{table:result} shows a summary of effectiveness values in terms of MAP for our model and the different baselines. \\  
Comparing different configurations of our approach, namely $DSRIM^{p2v}$, $DSRIM^{kr}$, and $DSRIM^{kr+p2v}$,  we can see that the DSRIM model applied only on our knowledge resource-driven representation $x^{KR}$ provides significant better performance (p-value$<$0.001) according to the MAP metric than the one with only the plain text-based representation $x^{t}$ (e.g., respectively $0.0307$ and $0.0183$ for the PMC dataset). This result reinforces the intuition claimed in recent work dealing with the use of text representations based on local interactions of terms and/or non-learned features \cite{guo2016deep}. Moreover, when combining the distributional and the relational semantics through the $DSRIM^{kr+p2v}$ model, we could see that the MAP value slightly increases, with for instance a significant improvement of $+67.09\%$ and $+87.98\%$ for the GOV2 and PMC datasets respectively with respect to $DSRIM^{p2v}$. This opens interesting perspectives in the combination of those word-sense approaches as we claim in this paper. 

With this in mind, we comment the baseline results with respect to the $DSRIM^{kr+p2v}$ model.
From a general point of view, we can see, on the one hand, that exact matching models are  non-significantly different from our proposed model, with a particular attention to the GOV2 dataset with small improvements with respect to BM25 ($+4.84\%$) and LM-DI ($+17.61\%$). On the other hand, our approach overpasses semantic and deep matching models with  significant improvements. For instance, our model reports significant better results for the GOV2 dataset according to the MAP  compared with the LM-QE, LM-LDA, DSSM, and CLSM models for which our model obtains a MAP value up to $+410.41\%$ of improvement rate. Those observations are similar for both datasets, highlighting the fact that our model is effective for leveraging  general (WordNet) as well as domain-oriented (MeSH) knowledge resources. More particularly, we can formulate the following statements:

%\begin{figure*}[t]
%\centering
%\includegraphics[width=0.7\textwidth,keepaspectratio]{images/in_out_tsne.png}
%\vspace{-0.2cm}
%\caption{Comparative analysis of the t-SNE projection between document and query input vectors (a)(c) and output vectors (d)(d) for query 789 of GOV2 dataset. $\circ$: relevant documents - $\bullet$: irrelevant documents.}
%\vspace{-0.2cm}
%\label{fig:inoutvectors}
%\end{figure*} 

\noindent $\bullet$ The BM25 and the language models are well-known as strong IR baselines which are difficult to outperform with deep matching models learned with small training datasets that do not allow to generalize the task. The results presented in Table 3 lead us to confirm this statement. However, it is worth noting that, in contrast to most previous neural approaches based on siamese architecture \cite{huang:dssm,severyn:shorttextrank,shen:clsm}  that rank  short documents (titles) and use large-scale real collection for training their model, we rather experiment our model on long full-text document collections  (average length is $1132.8$ words for  GOV2 and $477.1$ for  PMC). 
To get a better understanding of these results, we investigate to what extent the effectiveness of our model depends on the level of difficulty of queries. More particularly, we classify  queries according to three levels of difficulty (``easy'', ``medium'', ``difficult'') using the k-means algorithm applied on the BM25 MAP values. Statistics of each class are presented in Table \ref{tab:queries}. We can outline that, for the PMC dataset, difficult queries significantly include more terms and more objects than easy and medium ones. However, there is no significant differences between the different query types with respect to the number of terms and objects for the GOV2 dataset. 
Focusing on the retrieval effectiveness, it can be seen that   $DSRIM^{kr+p2v}$ improvements according to  BM25 are both positive and significant for difficult queries for both GOV2 and PMC. Moreover, it is worth mentioning that the improvement rates for difficult queries ($+63.60\%$ for the PMC dataset) are significantly different from  the ones for medium and easy queries (respectively $-25.78\%$ and $-0.22\%$ for the PMC dataset, with no significant improvement difference between easy and medium queries, $p>0.5$). Interestingly, combining the improvement rates and the number of objects for medium queries of the GOV2 dataset, we can see that the significant effectiveness decrease of our model ($-5.15\%$) could be explained by the lowest number of objects associated with this query set.
These results highlight that leveraging the relational semantics through our knowledge resource-driven representation is more effective for solving difficult queries. This is coherent since those queries are generally characterized by a high number of words and extracted objects. Accordingly, we can reasonably argue that our model is particularly devoted to lowering the semantic gap between word-based and concept-based representations of documents and queries which probably favors the discrimination between relevant and irrelevant documents.

\begin{table}[t]
\centering
%\vspace{-0.4cm}
\caption{Statistics on queries w.r.t their difficulty level}
\vspace{-0.4cm}
\begin{tabular}{cc|ccc}
\hline
 &Difficulty level& \#Words & \#Objects & \%Change   \\
 \hline
 \multirow{3}{*}{GOV2}&Easy & 22.95 & 12.11 & -16.60\% \\
 &Medium & 20.79 & 11.79 & -5.15\%*  \\
 &Difficult  & 22.15 & 12.14 & +87.15\%***  \\
 %\cline{2-5}
 %& Diff. vs. Med./Easy & &  & *\\
 \hline
  \multirow{3}{*}{PMC}&Easy & 13 & 5.4 & -0.22\%  \\
 &Medium & 16.68 & 5.36 & -25.78\% \\
 &Difficult & 18.5 &  6.3 & +63.60\%* \\
 %\cline{2-5}
 %& Diff. vs. Med./Easy & **& * & ***\\
 \hline
\end{tabular}
\vspace{-0.5cm}
\label{tab:queries}
\end{table}

\noindent $\bullet$ The LM-QE baseline performs a knowledge resource-based query expansion. Since the DSRIM outperforms the LM-QE model, we can suggest that the semantic based representations of documents and queries which are learned starting from the input built upon the relation mapping method, is more effective than the expanded queries with relevant object descriptors. \\
\noindent $\bullet$ The LDA-LM model is based on a probabilistic generative model able to identify relevant topics. Our model generally outperforms this baseline with a significant improvement of $89.95\%$ for the MAP metric on the PMC dataset. This is consistent with previous work \cite{huang:dssm}, highlighting the effectiveness of deep latent representations of texts in comparison to those obtained by generative models. \\
$\bullet$ In the category of neural IR models, our model outperforms the DSSM and the CLSM models (with a MAP reaching $0.0418$ and $0.0095$ for both datasets respectively). These results suggest that the integration of relational as well as the distributional semantics at the document level (rather than the word level) into the input representation allows enhancing the learning of the deep neural matching model while considering small collections (instead of real search logs) and full texts (instead of titles). 
Interestingly, the convolutional CLSM model initially overpassing the DSSM in \cite{shen:clsm} through experiments carried out on a large-scale real-world data, is less effective than the DSSM. One explanation might be that it is trained using TREC collections characterized by a limited number of queries (as also shown in \cite{guo2016deep}). 
A further analysis based on the cosine similarity between document-query vectors of input and output relevant pairs  obtained using both  DSSM and  DSRIM  highlights that the use of evidence from relational semantics underlying queries and documents allows a better discrimination between relevant and irrelevant documents. Indeed, the similarity improvement between input/output representations is more important for our model than for the DSSM model for both datasets: $166.88\%$ for DSSM vs. $271.51\%$ for DSRIM for the GOV2 dataset, $5.91\%$ for DSSM vs. $71.71\%$ for the PMC dataset.

%In order to further investigate the impact of incorporating evidences issued from the external knowledge in a deep model, we report in Table \ref{tab:sim} the measures of the cosine similarity between document-query vectors of input and output relevant pairs  obtained using both  DSSM and  DSRIM. As can be seen from Table \ref{tab:sim}, although input similarities are of the same range for both datasets, the similarity improvement between input/output representations is more important for our model than for the DSSM model for both datasets: $166.88\%$ for DSSM vs. $271.51\%$ for DSRIM for the GOV2 dataset, $5.91\%$ for DSSM vs. $71.71\%$ for the PMC dataset.  
\begin{comment}
\begin{table}[t]
\centering
%\vspace{-0.2cm}
\caption{Average similarity of relevant document-query pairs at the I/O layers }
\vspace{-0.4cm}
\label{tab:sim}
\begin{tabular}{cll|ll}
\hline 
 & \multicolumn{2}{c|}{GOV2} & \multicolumn{2}{c}{PMC} \\
 & Input 	&Output& Input & Output\\
 \hline
DSSM~~ &	0.1482~~ 	& 0.3955~~ &0.1049 &0.1111  \\
DSRIM & 	0.1988 	& 0.7386 &0.3659 & 0.6283\\
\hline
\end{tabular}
\vspace{-0.2cm}
\end{table}
\end{comment}

% ======================
% Section 6: Conclusion
% ======================
\section{Conclusion}
We propose the DSRIM model, a deep neural IR model that leverages both distributional semantics through the \textit{ParagraphVector} algorithm, and relational semantics, through a knowledge resource-driven representation of texts aiming at jointly modeling  embedded objects and structured relations between objects. 
Experimental evaluation on two TREC datasets, namely the GOV2 and the PMC Open Access, are performed to evaluate the quality of the input representations as well as their impact on document ranking effectiveness. Results show that 1) our knowledge resource-driven representation allows to discriminate semantically similar from non-semantically similar texts, and that 2) our model overpasses semantic-driven approaches as well as state-of-the-art neural IR models.
In the near future, we plan to further the knowledge resource-driven representation by taking into account both the heterogeneity of objects and the heterogeneity of the relations between objects. 

\section*{Acknowledgement}
This research was supported by the French FUI research program SparkInData.

\bibliographystyle{ACM-Reference-Format}
\bibliography{ictir} 

\end{sloppypar}
\end{document}